# $Ir_3Ge_{20}^-$: A $3n$-Connected Cloverleaf-Shaped Supercluster

Xue Wu,[1]  Wen-Shuai Dai,[2]  Lulu Li,[1]  Fangying Hao,[1]  Hong-Guang Xu,[2]\*  Wei-Jun Zheng,[2]  Jijun Zhao[3]\*

[1] State Key Laboratory of Metastable Materials Science and Technology & Hebei Key Laboratory of Microstructural Material Physics, School of Science, Yanshan University, Qinhuangdao 066004, China

[2] Beijing National Laboratory for Molecular Sciences, State Key Laboratory of Molecular Reaction Dynamics, Institute of Chemistry, Chinese Academy of Sciences, Beijing 100190, China

[3] Guangdong Basic Research Center of Excellence for Structure and Fundamental Interactions of Matter, Guangdong Provincial Key Laboratory of Quantum Engineering and Quantum Materials, School of Physics, South China Normal University, Guangzhou 510006, China

\* Author to whom any correspondence should be addressed.

---

\* Corresponding author. Email: xuhong@iccas.ac.cn; zhaojj@scnu.edu.cn

**Abstract**

Group 14 Zintl clusters are promising molecular building blocks for nanoscale architecture. Endohedral variants, which encapsulate *d*/*f*-block metals within *p*-block (semi)metal cages, provide insights into intermetallic bonding and compound formation. In this study, experimental photoelectron spectroscopy and first-principles calculations were used to investigate the Ir-doped germanium cluster species. A new $C_{2v}$ symmetric building block, IrGe$_{12}^-$, was identified, serving as the basis for designing the supercluster Ir$_3$Ge$_{20}^-$ with a cloverleaf-shaped, $D_{3h}$ symmetric architecture. This structure consists of three interconnected aromatic [IrGe$_{12}$] units linked by Ge–Ge σ bonds, forming shielding cones. Ir$_3$Ge$_{20}^-$ follows the 5*n* rule with 100 valence electrons, featuring a core-Ir$_3$ unit with a $d^{10}$ closed-shell configuration sharing electrons with the Ge$_{20}$ skeleton. The stability, chemical bonding, and aromaticity of Ir$_3$Ge$_{20}^-$ were confirmed, demonstrating a novel approach to precise atom manipulation in cluster-based materials and devices.

**Introduction**

The discovery and bulk synthesis of fullerenes has led to new research fields in chemistry and nanoscience, sparking significant interest in the quest for stable clusters.[1,2] Rapid advancements in fullerene-related clusters and the extensive applications of fullerene-based materials have spurred investigations into analogous hollow sphere molecules composed of other main groups or transition metal elements, termed inorganic fullerenes.[3] In contrast to carbon atoms, the 4$s$ orbitals of germanium atoms exhibit limited hybridization with the 4$p$ orbitals, resulting in the localization of $s$-type lone pairs with high occupancy.[4] Consequently, electron deficiency prevents the formation of stable symmetrical fullerene-like structures. However, doping with metals can supply additional electrons to compensate for the deficiencies of Ge atoms. Recent efforts have been dedicated to designing and synthesizing homologous fullerene cages, leading to the successful creation of homogroup fullerene cages such as metal-encapsulating $Ge_{16}$ cages.[5]

The 12-vertex clusters play a crucial role in supramolecular cluster assembly, exhibiting various unique geometries such as deltahedral structures, $D_{2d}$-symmetric 3-connected structures, and hexagonal prisms ($D_{6h}$). The geometry of these clusters is closely associated with the valence electron (VE) count in the $p$-block element shell. For instance, the deltahedral structure is demonstrated by 50-electron icosahedral clusters such as $[M@Pb_{12}]^{2-}$ (M = Ni, Pd, Pt)[6,7] and $[Ir@Sn_{12}]^{3-}$,[8] which exhibit nearly perfect icosahedral ($I_h$) symmetry. The bonding within these clusters is characterized by electron deficiency (< 2e$^-$/edge) and high delocalization, resembling $d^{10}$ metal atoms/ions enclosed within deltahedral cages (4$n$ + 2 = 50 electrons). However, anomalies exist, such as the anion $[Mn@Pb_{12}]^{3-}$, which formally contains 53 valence electrons in its cluster shell,[9] leading to distortion from $I_h$ to $D_{2h}$ owing to the partial transfer of valence electrons from the Mn atom. A similar example is the $D_{5d}$ $CoGe_{12}^{3-}$ cluster.[10] As the valence electron count increased, clusters with formally electron-precise (2e$^-$, 5$n$) cage structures emerged, such as $D_{2d}$-symmetric $[Ta@Ge_8As_4]^{3-}$ (VE = 60),[11] $M@Ge_{12}$ (M = Ni, Au),[12,13] $[Ru@Ge_{12}]^{3-}$,[14] hexagonal prism ($D_{6h}$) $W@Ge_{12}$[15] and $Cr@Si_{12}$.[16] As a result, the 60-valence electron (5$n$, $n$ = 12) cluster adopts a structure with only three connected atoms, fully eliminating deltahedral faces.

Transition metal derivatives of polyatomic main-group anions (Zintl ions) play a crucial role in bridging the chemistry of large clusters and nanoparticles.[17,18] Synthesized based on closo-deltahedral structures, multi-core intermetalloid nanoclusters, such as [Pd$_2$@Ge$_{18}$]$^{4-}$ exhibit a deltahedral capsule-like structure formed by merging the parent spherical aromatic [Pd@Ge$_{12}$]$^{2-}$ building units.[19,20] These nanoclusters display a highly dynamic behavior in solution, resembling a liquid with intramolecular exchange at all sites. Additionally, a distinct $D_{2h}$ 3-connected architecture, exemplified by bimetallic-doped [Co$_2$@Ge$_{16}$]$^{4-}$ clusters, can be viewed as the face fusion of two [Co@Ge$_{10}$] units through a Ge$_4$ face.[21] Furthermore, double-cage Zintl clusters can be synthesized using different linking methods, such as shared edges or atom connections. For instance, the edge-sharing endohedral Ge cluster [Co$_2$@(Ge$_{17}$Ni)]$^{4-}$ was fused through two [Co@(Ge$_9$Ni)] moieties by Ni insertion, resulting in a new atomic arrangement.[22] Another example is the synthesis of the sandwich-type anion [NiGe$_9$-In-NiGe$_9$]$^{5-}$, which shows the reactivity of sandwich-type [Ge$_9$-In-Ge$_9$]$^{5-}$ as a precursor.[23] Moreover, the synthesis and characterization of two 17-atom Co-filled Ge cluster dimers, [NiGe$_9$-Ni-NiGe$_9$]$^{4-}$, demonstrated a true local minimum, as confirmed by quantum chemical calculations.[24]

In addition, a triple-cage cluster of [Rh$_3$@Sn$_{24}$]$^{5-}$ consists of three face-fused Rh@Sn$_{10}$ units arranged in an approximately $C_{3v}$-symmetric geometry, with the trimer sharing a single square face of Rh@Sn$_{10}$.[25] Another triple-cage cluster, [Pd$_3$Sn$_8$Bi$_6$]$^{4-}$, comprises three Pd@Sn$_6$Bi$_3$ units that share an Sn$_4$ rectangular face with each other.[26] Despite the relatively small cavity inside the cage, the Pd$_3$ triangle interacted weakly with the Sn/Bi cluster shell. Although the Pd$_3$ cluster has a weak interaction with the main-group metal shell, it is best described as neutral Pd$_3$ trapped within the comparatively small cavity of the [Sn$_8$Bi$_6$]$^{4-}$ anion.

Despite extensive research on multi-core intermetalloid nanoclusters, the electronic structure of these clusters remains unclear, impeding the application of electron counts in nanoclusters. To address this knowledge gap, we conducted a comprehensive theoretical and experimental investigation of the Ir-doped Ge clusters. A metastable state, Ir$_1$Ge$_{12}^-$, was proposed by combining a closo-deltahedral and 3-connected structure with superatom orbitals. Based on this metastable state, an Ir$_3$Ge$_{20}^-$ supercluster was designed, featuring a cloverleaf structure with $D_{3h}$ symmetry. The Ge$_{20}$-skeleton adheres to the 5$n$ electronic counting rule (100$e$) and shares 22 electrons with core-Ir$_3$, resulting in a $d^{10}$ closed-shell configuration that maximizes electrostatic interactions between the metal and the shell. This study expands the 12-vertex Ge cluster family and lays a theoretical and practical foundation for the design of multicentered intermetalloids.

**Method**

**Experimental**

The experiments were performed using a magnetic-bottle photoelectron apparatus equipped with a laser vaporization cluster source, the details of which have been described previously.[27] Briefly, Ir-doped germanium clusters were produced in the laser vaporization source by laser ablation of a rotating and translating disk target of a mixture of iridium and isotopically enriched germanium (13 mm diameter, Ir: $^{74}$Ge mole ratio 1:16, $^{74}$Ge 99.80%) with second harmonic (532 nm) light pulses from a Nd: YAG laser (Continuum Surelite II-10). Helium gas with ~4 atm backing pressure was allowed to expand through a pulsed valve (General Valve Series 9) into the source to cool the formed clusters. Ir-doped germanium clusters were extracted from the cluster beam perpendicularly, and the mass was analyzed using a time-of-flight mass spectrometer. $Ir_1Ge_{12}^-$, $Ir_2Ge_{17}^-$ and $Ir_3Ge_{20}^-$ were mass-selected and decelerated before being photodetached by a 266 nm laser beam from another Nd: YAG laser (Continuum Surelite II-10). Photoelectrons were analyzed using a magnetic bottle photoelectron spectrometer. The magnetic-bottle photoelectron spectrometer had an energy resolution of $\Delta E/E \approx 4.0\%$, corresponding to approximately 40 meV for 1 eV kinetic energy electrons. The photoelectron spectra were calibrated using the spectra of the $Pb^-$, $Bi^-$ and $I^-$ ions obtained under similar conditions.

**Theoretical**

An unbiased global search for the global minima of $Ir_1Ge_{12}^-$, $Ir_2Ge_{17}^-$ and $Ir_3Ge_{20}^-$ were performed using a comprehensive genetic algorithm (CGA) code[28] combined with density functional theory (DFT) calculations.[29] Throughout the genetic algorithm iterations, the child clusters were relaxed using the Perdew-Burke-Enzerhof functional[30] and the double numerical polarization function basis, including *d*-polarization function (DND),[30,31] as incorporated in the DMol$^3$ program.[29] The reliability and efficiency of this CGA-DFT approach have been validated in prior studies on doped Group 14 Zintl ions.[32-36] More details on the CGA methodology can be found in a review article.[28]

The low-lying isomers identified through the CGA-DFT search were further optimized by exploring various spin multiplicities with high accuracy using the B3PW91 functional.[37] The optimization employed 6-311+G(d) and SDD basis sets for Ge and Ir atoms, respectively. The vibrational analysis confirmed the stability of the reported isomers by ensuring the absence of imaginary frequencies. The relative stabilities of the clusters were assessed based on the average binding energies per atom ($E_b$) and HOMO-LUMO gaps ($E_{gap}$). The calculations were performed using the Gaussian16 package.[38]

Chemical bonding was analyzed using the electron localization function (ELF)[39-41] and the Adaptive Natural Density Partitioning (AdNDP)[42] methods, while aromaticity was assessed through nucleus independent chemical shifts (NICS).[43,44] The atomic dipole moment-corrected Hirshfeld (ADCH) method was employed to determine the atomic charge distributions.[45] The evaluation of chemical bonding and aromaticity were performed using the Multiwfn 3.8(dev) code.[46]

**Results and Discussion**

The icosahedron cage, a well-defined structural unit found in various bulk materials, is renowned for its stability and rigidity and plays a pivotal role in enhancing the overall strength and stability of these materials.[47] Starting with Ir-doped $Ge_{12}$ species, we observed that the $Ge_{12}$ icosahedral cage was insufficient to host the Ir atom, resulting in the deformation of the icosahedral cage with $D_{5d}$-symmetry. Consequently, an extensive search for the global minima of Ir-doped $Ge_{12}$ species was conducted using a comprehensive genetic algorithm (CGA) code combined with DFT calculations. The results show that there are two open-cage configurations with relatively low symmetry and a lower energy preference (Figure 1). The well-known 3-connected $D_{2d}$-symmetric bicapped pentagonal prism (BPP)[14] ($Ir_1Ge_{12}^-$-I) and a newly discovered $C_{2v}$ 12-vertex isomer, denoted as $Ir_1Ge_{12}^-$-II, features a $Ge_2$ dimer attached to the waist of a pentagonal prism. This unique isomer exhibits a combination of a half 3-connected structure and half-icosahedral structure, thereby expanding the 12-vertex family.

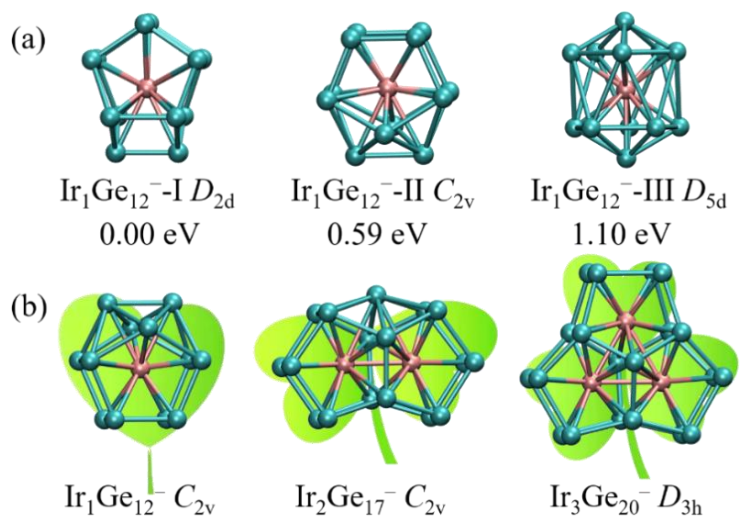

Figure 1. (a) The low-energy structures of the $Ir_1Ge_{12}^-$ clusters with the cluster symmetry and relative energy (in eV), (b) the trefoil $Ir_1Ge_{12}^-$ unit block, $Ir_2Ge_{17}^-$ and $Ir_3Ge_{20}^-$ clusters. Blue/green and pink balls represent Ge and Ir atoms, respectively.

The stability of the 3-connected BPP ground state structure is due to the $5n$ electron precise count at the cage, with all five $d$ orbitals of the electron-rich metal atom participating in back-bonding.[14,48] The $Ir_1Ge_{12}^-$-I BPP ground state, with 58 valence electrons (exactly $5n - 2 = 58$) is particularly stable. One reason for this is that the 58 valence electrons form a magic number configuration, with an electron occupation of $1S^2 1P^6 1D^{10} 1F^{14} 2S^2 2P^6 2D^{10} 1G^8$, as shown in Figure S1 (a). Another reason is that the LUMO is localized symmetrically on the $Ge_{12}$ cage, which means that the removal of two electrons does not directly compromise the Ir–Ge bonding. Instead, only the bonds within the $Ge_{12}$ framework were weakened. This is similar to the 3-connected $RuGe_{12}^{3-}$ cluster, with 59 valence electrons.[14] The metal centers in the 58-electron clusters have lower effective nuclear charges than their 60-electron counterparts, which enhances Ir–Ge bonding (the $d$ orbitals are more valence-like). This compensates for the absence of two (weak) Ge–Ge bonding electrons. The metastable isomer $Ir_1Ge_{12}^-$-II, with the same electron configuration as shown in Figure S1 (b) and adhering to the 58-electron magic number rule, exhibits a HOMO-LUMO gap very close to that of the ground state (1.39 eV compared to 1.67 eV). From an electronic structure perspective, the key distinction lies in the relatively delocalized LUMO over the Ir–Ge bonds, resulting in a deficiency of two electrons, which weakens the Ir–Ge bonding compared to the ideal 60 electrons. This may explain the higher energy (0.59 eV) compared to that of the ground state.

Within the domain of supramolecular cluster assembly, our goal was to devise a multicentered intermetalloid by employing 12-vertex isomers. The fundamental principle is to guarantee the engagement of every Ge atom in a three-coordinate supramolecular configuration following the $5n$ rule. Subsequently, a $D_{3h}$ three-centered intermetalloid was effectively synthesized using the metastable $Ir_1Ge_{12}^-$-II complex, as shown in Figure 1(b). This multicentered intermetalloid trimer was constructed from three building blocks, creating a tri-supramolecular structure comprising three pentagonal face-sharing $Ir_1Ge_{12}$ units. These units feature tightly interwoven Ir cores occupying Ge vertices, resulting in minimal structural relaxation. Consequently, the $D_{3h}$ $Ir_3Ge_{20}^-$ cluster can be conceptualized as an amalgamation of three endohedral germanium spherenes along the $C_2$ axis. Notably, computational global conformational analysis indicated that $D_{3h}$ $Ir_3Ge_{20}^-$ is the most stable ground state of the anionic complex, with an energy 1.01 eV lower than that of the second lowest-lying isomer $C_s$ $Ir_3Ge_{20}^-$-II at the B3PW91/6-311+G(d) level (refer to Figure S2 in the SI and Table 1, respectively).

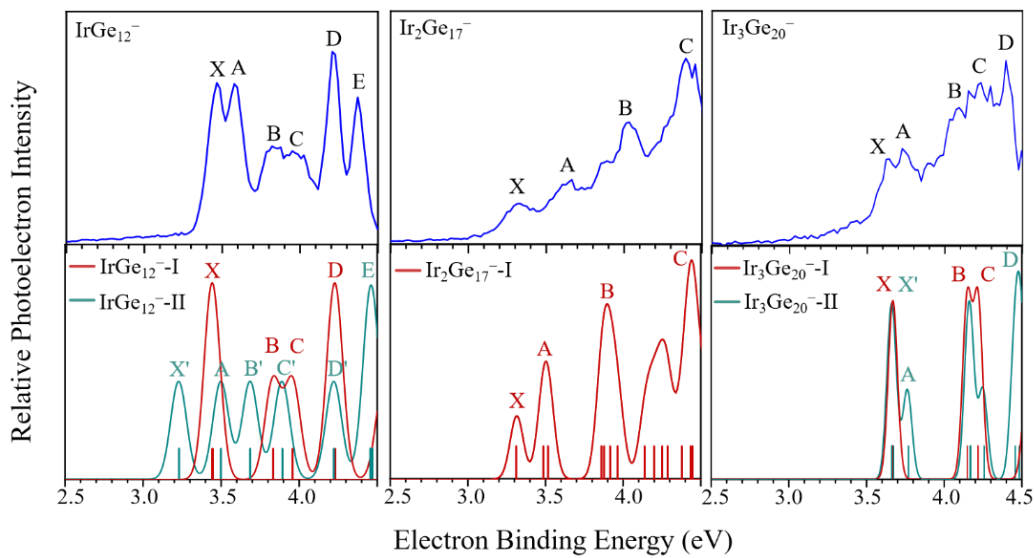

Figure 2. Photoelectron spectra of $Ir_1Ge_{12}^-$, $Ir_2Ge_{17}^-$ and $Ir_3Ge_{20}^-$ clusters recorded with 266 nm photons and photoelectron spectra of DFT-calculated clusters. Blue lines represent experimental while red and green represent theorical photoelectron spectra, respectively.

The trefoil $Ir_1Ge_{12}^-$, $Ir_2Ge_{17}^-$ and $Ir_3Ge_{20}^-$ clusters were confirmed by experimental photoelectron spectroscopy (PES) using 266 nm photons, as shown in Figure 2 (blue lines). The PES of $Ir_3Ge_{20}^-$ indicated two distinct bands (X and A) in the low-binding-energy range and nearly continuous signals between 4.0 and 4.5 eV. Band X yielded a ground state vertical detachment energy (VDE) of 3.54 eV, while the adjacent band A yielded a VDE of 3.75 eV. Subsequently, a region of continuous higher binding energy signals (labeled B, C, and D for discussion purposes) was observed, likely arising from multiple detachment transitions in the large system. Despite the spectral complexity, the well-defined X and A bands, along with the energy gap, present crucial spectral features for comparison with the theoretical predictions.

To verify the global minimum of $Ir_3Ge_{20}^-$, the VDEs of the two lowest-lying isomers were computed. The computed VDEs were convolved with unit-area Gaussians of 0.07 eV width to produce a simulated photoelectron spectrum, which were then compared with experimental data at 266 nm in Figure 2 (red and green lines). The VDE values for the $Ir_3Ge_{20}^-$ clusters are compared with the experimental results presented in Table 1. The comparison reveals a strong agreement between the simulated spectral patterns of the global minimum and metastable isomer of $Ir_3Ge_{20}^-$ with the experimental data. The VDEs of the two isomers, $Ir_3Ge_{20}^-$-I and $Ir_3Ge_{20}^-$-II, were determined to be 3.67 and 3.66 eV, respectively, consistent with the experimental VDE of 3.64 eV for the X band. The A band originated exclusively from the second VDE of metastable $Ir_3Ge_{20}^-$-II, separated by an energy gap of approximately 0.4 eV. The higher binding-energy transitions from $Ir_3Ge_{20}^-$-I and $Ir_3Ge_{20}^-$-II corresponded well with the observed spectral bands B−D, each exhibiting multiple detachment channels. This observation provides supports the presence of coexisting $Ir_3Ge_{20}^-$-I and $Ir_3Ge_{20}^-$-II isomers as the underlying cause of the observed photoelectron spectrum of $Ir_3Ge_{20}^-$.

Table 1. Theoretical and experimental vertical detachment energy (VDE in eV), energy difference ($\Delta E$ in eV), binding energy ($E_b$ in eV per atom), HOMO-LUMO gap ($E_{gap}$ in eV), Mulliken population analysis and Bader charges of Ir atom (in $e$ per atom) for $Ir_1Ge_{12}^-$, $Ir_2Ge_{17}^-$ and $Ir_3Ge_{20}^-$ clusters.

| Clusters | Sym. | $\Delta E$ | $E_b$ | $E_{gap}$ | VDE Expt. | VDE Theo. | Mulliken | Bader |
|---|---|---|---|---|---|---|---|---|
| $Ir_1Ge_{12}^-$-I | $D_{2d}$ | 0.00 | 4.70 | 1.67 | | 3.44 | −1.54 | −1.09 |
| $Ir_1Ge_{12}^-$-II | $C_{2v}$ | 0.59 | 4.65 | 1.39 | 3.46 | 3.23 | −1.45 | −1.28 |
| $Ir_1Ge_{12}^-$-III | $D_{5h}$ | 1.10 | 4.61 | 1.57 | | 3.48 | −1.06 | −1.14 |
| $Ir_2Ge_{17}^-$-I | $C_{2v}$ | 0.00 | 4.77 | 2.03 | 3.33 | 3.32 | −0.94 | −1.03 |
| $Ir_3Ge_{20}^-$-I | $D_{3h}$ | 0.00 | 4.88 | 1.89 | 3.64 | 3.67 | −0.84 | −0.97 |
| $Ir_3Ge_{20}^-$-II | $C_s$ | 1.01 | 4.84 | 1.78 | | 3.66 | −0.84 | −0.92 |

The stability of the ground-state $D_{3h}$ $Ir_3Ge_{20}^-$-I configuration with 20 vertices can be explained by the 5$n$ valence electron rule, which is a consequence of the nature of the three-coordinated main group atom. This configuration was achieved when considering 80 electrons from the 20 Ge atoms and 1 from the charge, along with the 19 $d$ electrons on $Ir_3$, totaling 100 electrons. Meanwhile, Ir tends to maintain all of its valence electrons in the $d^{10}$ shell for chemical feasibility. Consequently, the $Ge_{20}$-skeleton shares 22 electrons with core-$Ir_3$, maximizing the electrostatic interactions between the metal and the shell (illustrated in a Venn diagram in Scheme 1). The $Ge_{20}$-skeleton adheres to the 5$n$ valence electron rule, while core-$Ir_3$ possesses 30 valence electrons (22 delocalized electrons + 8 localized electrons) forming a $d^{10}$ closed-shell, which was validated by the Kohn-Sham molecular orbitals (MOs) shown in Figure 3. The results revealed four nonbonding orbitals primarily on core-Ir (pink energy levels) and 11 doubly occupied orbitals on the core-Ir and Ge-skeletons with significant bonding characteristics (blue energy levels). Furthermore, the frontier molecular orbitals of the doubly occupied orbitals play a crucial role in enhancing the Ir-Ge bonding, thereby contributing to the overall stability of the cluster. Additionally, the $Ge_{20}$-skeleton transfers two extra electrons to the $Ir_3$-core, as confirmed by the natural orbitals for chemical valence (NOCV) analysis[49] depicted in Figure S3 of the SI. Mulliken population analysis and Bader charge analysis further indicated that each Ir atom carried approximately one negative charge. These findings highlight the electronic interactions and bonding properties of the clusters.

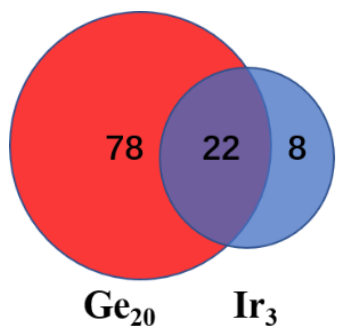

Scheme 1. Venn diagram representation of the Jigsaw electron counting scheme of $Ir_3Ge_{20}^-$-I cluster.

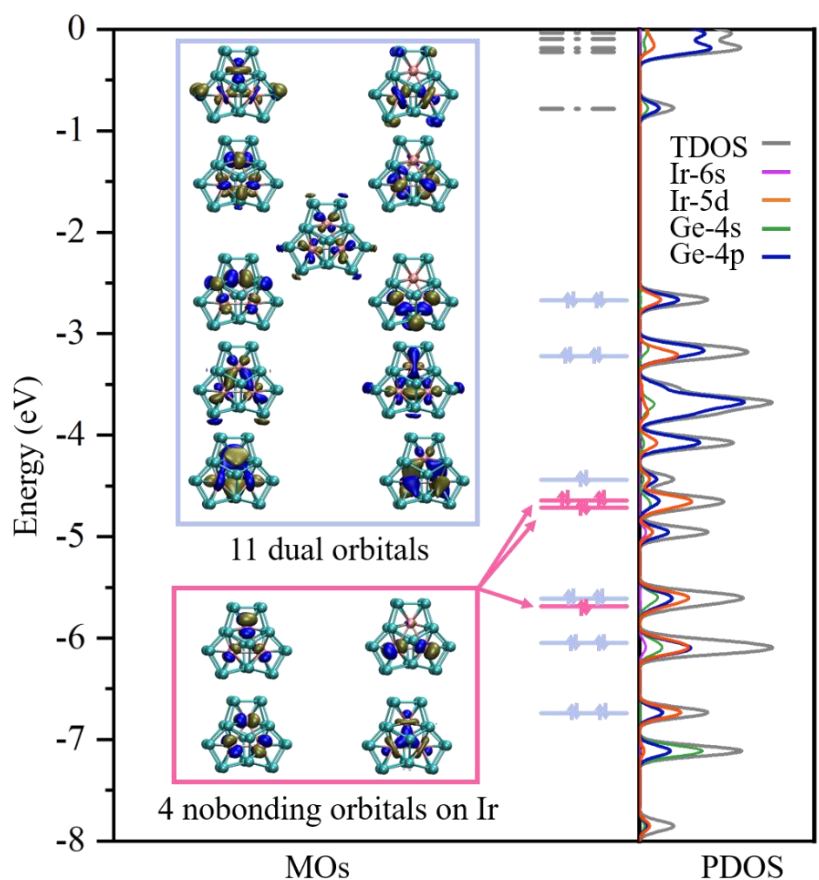

Figure 3. Frontier Kohn-Sham orbitals and molecular orbital (MOs) diagram and partial density of states (PDOS) of the anionic $Ir_3Ge_{20}^-$-I cluster. Orbitals with both core-Ir and Ge-skeleton contribution are shown in left column, corresponding to blue energy levels; orbitals with dominant Ir contribution are shown in right column, corresponding to pink energy levels.

The structural stability of the optimized anionic $Ir_3Ge_{20}^-$-I cluster was investigated using the AdNDP bonding analysis (Figure 4). This revealed 15 $d$ lone pair electrons on the three Ir atoms, with three $d_{xz}$ lone pair electrons forming in-plane 2c−2e σ bonds with ON = 1.87 |e|, representing a 30-valence electron $d^{10}$ closed-shell occupation. The $d_{x2-y2}$ and $d_{z2}$ lone pair electrons had ON values exceeding 1.90 |e|, while $d_{yz}$, $d_{xy}$ and $d_{xz}$ lone pair electrons had lower ON values, indicating shared electrons with the $Ge_{20}$-skeleton. The peripheral Ge atoms carried 12 $s$ lone pair electrons, six 2c−2e σ bonds on the top and bottom Ge-Ge bonds, 12 2c−2e σ bonds on the peripheral Ge-Ge bonds, and three 4c−2e σ bonds and three 5c−2e σ bonds on each trefoil leaf. This arrangement was supported by the ELFσ results (Figure 4a). Additionally, three electron pairs participated in two delocalized σ bonds and a π-bonding system spanning the entire trefoil structure, which was consistent with the ELFπ results showing complete electronic delocalization. The $Ir_3Ge_{20}^-$-I cluster displayed π-aromaticity, satisfying the criteria for $2(N+1)^2$ (N = 0) near-spherical aromaticity. The bonding pattern aligns with the in-plane color-filled map shown in Figure 4b. A high ELF distribution (close to 1) at the peripheral Ge and inner Ir atoms signifies lone pair electrons, whereas strong covalent bonding is evident in the peripheral Ge-Ge bonds. Conversely, interactions involving Ir atoms are weak and exhibit ionic bond characteristics.

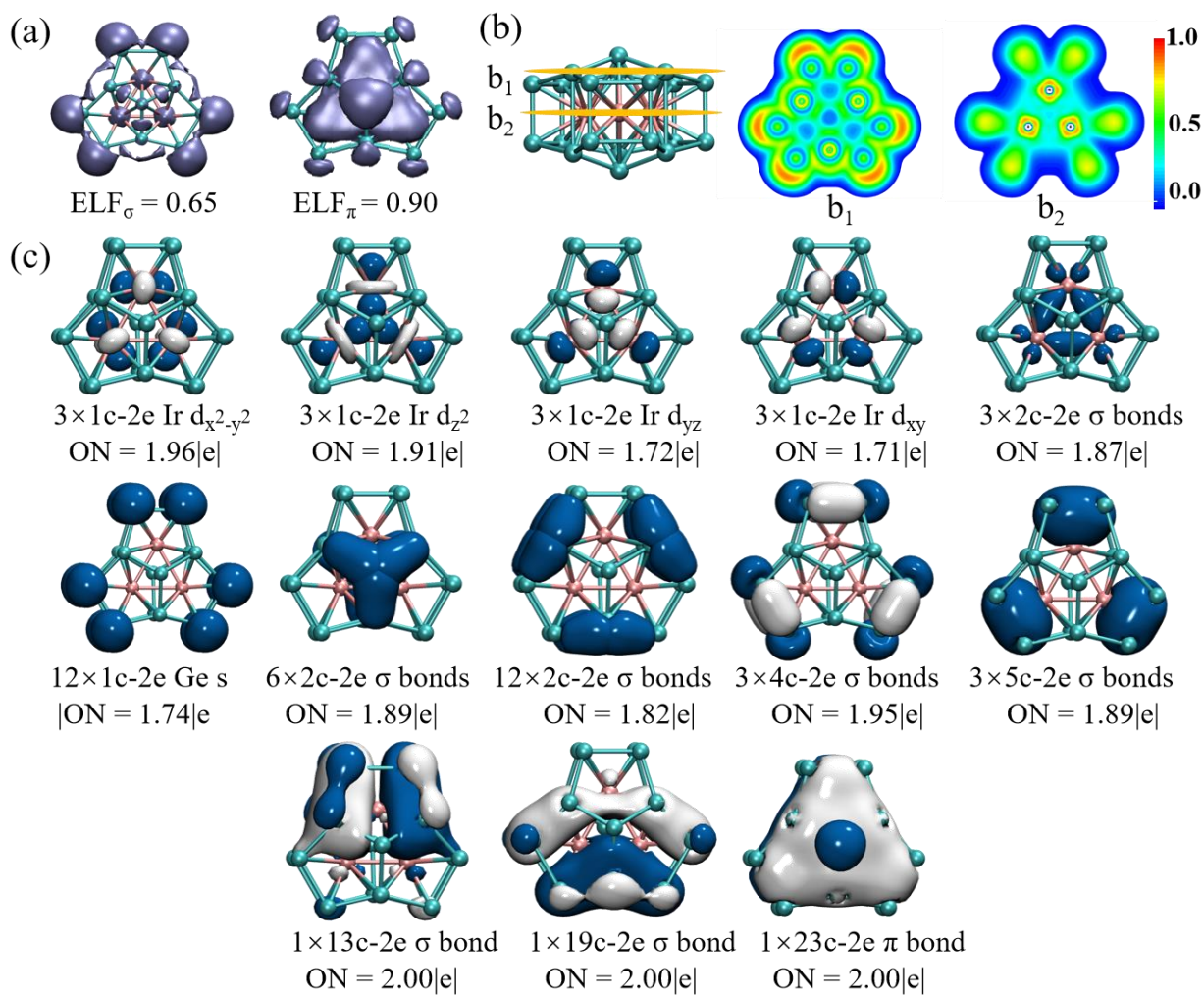

Figure 4. (a) ELF analysis of the anionic $Ir_3Ge_{20}^-$ cluster and (b) total ELF color-filled map in the central XY plane $b_1$ across the $Ge_3$ plane of the top layer and $b_2$ across the $Ir_3$ plane, (c) AdNDP analysis of the anionic $Ir_3Ge_{20}^-$-I cluster. ON denotes the occupation number.

To further investigate the aromatic properties of the anionic $Ir_3Ge_{20}^-$ cluster, we conducted NICS(1) calculations. An NICS(1) color-filled plane map was generated at a distance of approximately 1 Å from the center of the cluster (Figure 5a), along with scanned curves of the NICS values in two directions (Figure 5b, c). Our findings reveal that the $Ir_3Ge_{20}^-$ cluster demonstrates typical aromaticity, with the core-Ir atoms exhibiting a negative NICS(1) value (approximately −46 ppm), and the $Ge_{20}$-skeleton also displayed negative NICS(1) values ranging from −60 to −75 ppm. The NICS-scan curves along the y- and z-axes indicate the three-dimensional aromatic nature of the system.

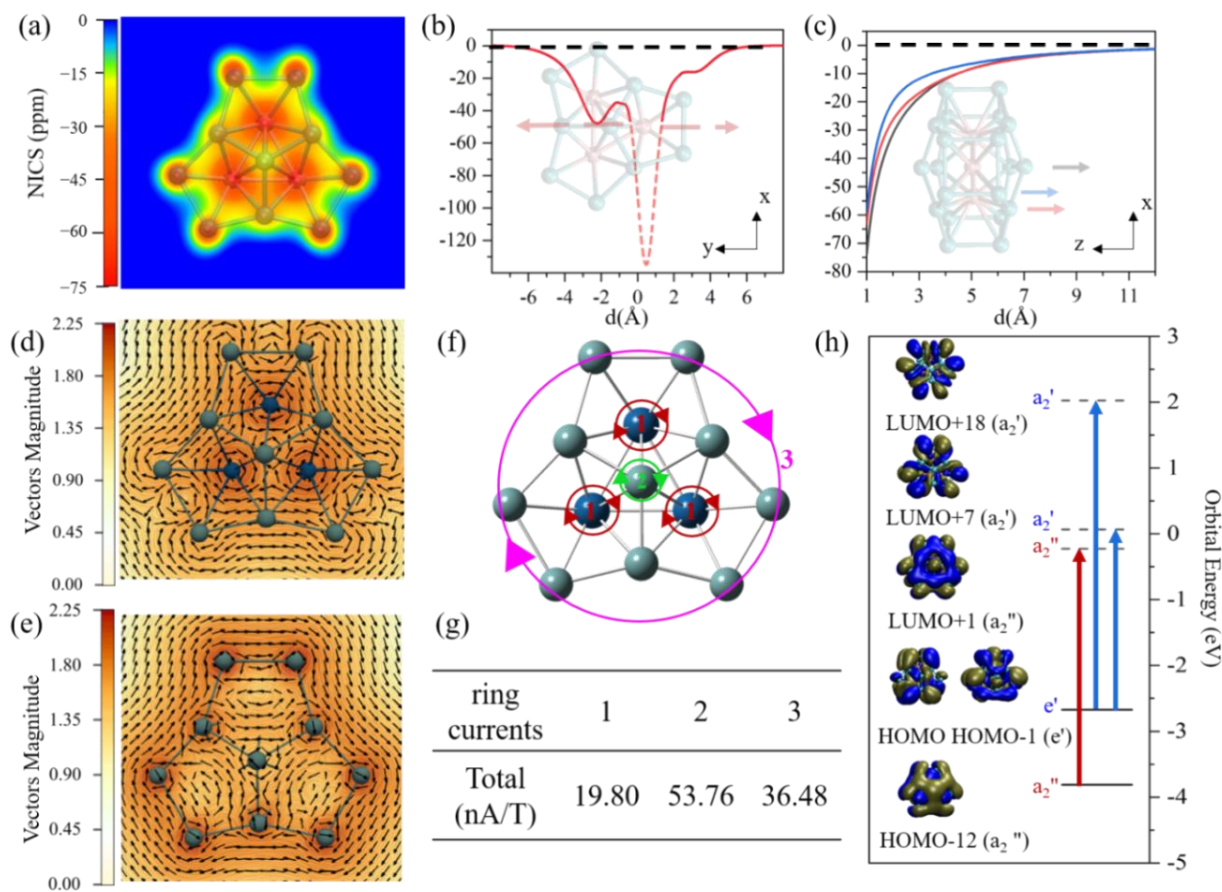

Figure 5. (a) NICS(1) color-filled plane map above the central XY plane for 1 Å of the $Ir_3Ge_{20}^-$ cluster. NICS-scanned curves on (b) the y-axis and (c) the z-axis. GIMIC map with vector plots of the calculated current density for (d) the central core-$Ir_3$ XY plane and (e) Ge layer XY plane. (f) Identified ring current circuits in the core-Ir, top-Ge, and $Ge_9$ fragment, with their respective ring current strength values in nA $T^{-1}$ in (g). The external magnetic field perpendicular to the ring plane and pointing outward, induces diatropic currents corresponding to clockwise electrons circulation. (h) Frontier orbital energy levels (in eV) with major translational transitions indicated by red (π) and blue (σ) full arrows.

The magnetic response of the $Ir_3Ge_{20}^-$ compound to an external magnetic field was assessed using a gauge-including magnetically induced current (GIMIC)[50,51] map (Figure 5d, e) to analyze the ring current strength profiles within the plane. The identified ring current circuits and their strengths derived from the analysis of the integrated current density profiles are depicted in Figure 5f and g. The isosurfaces of the current density plot reveal three diatropic rings: one encompassing the core-Ir atoms (19.80 nA $T^{-1}$), another around the top-Ge atom (53.76 nA T$^{-1}$) and a third surrounding the outer Ge$_9$ fragment of the cluster (36.48 nA T$^{-1}$). This significant magnitude, notably higher than the net ring current strength of 14.2 nA T$^{-1}$ for C$_6$H$_6$ at the same level, suggests an aromatic character of the Ir$_3$Ge$_{20}^-$ cluster.

Based on the previous analyses, the excitation responsible for the magnetic response was determined to elucidate the aromatic character (Figure 5h). According to the CTOCD-DZ method,[52] the diamagnetic current density is attributed to virtual translational transitions, whereas the induced current density is linked to the virtual excitation of a few electrons in the frontier orbitals. Specifically, for the Ir$_3$Ge$_{20}^-$ cluster, which contains six electrons in the singlet state, with a π electronic configuration of (a$_2$'')$^2$ and a σ electronic configuration of (e')$^4$, the diatropic current density in both the π- and σ-electron subsystems arises from the translational transition between HOMO−12 and LUMO+1 and from the degenerate HOMO (HOMO−1) to LUMO+7 and LUMO+18, respectively. However, paratropic currents resulting from rotational transitions with larger orbital energy differences, which could slightly offset the contribution of diamagnetic ring currents to aromaticity, are not shown in this figure. This analysis confirmed the presence of diamagnetic ring currents in both σ and π electrons, indicating enhanced diamagnetism due to valence delocalization and supporting the dual σ and π aromaticity of the Ir$_3$Ge$_{20}^-$ cluster.

**Conclusions**

In conclusion, we have successfully modeled the targeted synthesis of a three-centered intermetalloid, Ir$_3$Ge$_{20}^-$, using metastable $C_{2v}$-Ir$_1$Ge$_{12}^-$ as a building block that combines a half 3-connected structure with a half icosahedral structure, thus expanding the 12-vertex family. The Ir$_3$Ge$_{20}^-$ cluster, a trimer composed of three iridium-encapsulated germanium cage units fused together by sharing their pentagonal faces, exhibited $D_{3h}$ symmetry, which was confirmed to be the ground state through both experimental photoelectron spectroscopy and theoretical simulations. The $D_{3h}$-Ir$_3$Ge$_{20}^-$ cluster with 20 vertices follows the 5$n$ valence electron rule, where the Ge$_{20}$ skeleton contributes 80 electrons, the Ir$_3$ core provides 19 $d$ electrons, and the cluster charge adds one electron. Notably, the Ge$_{20}$ skeleton shares 22 electrons with Ir$_3$, enabling Ir to achieve a $d^{10}$ closed-shell configuration (30 valence electrons) and maximize Ir–Ge electrostatic interactions. Furthermore, this three-centered intermetalloid exhibits dual σ and π aromatic character, primarily because of the local contributions of the Ir$_3$ core and delocalized Ge$_{20}$ skeleton. This study elucidates the structural and electronic origins of the stability in Ir-doped Ge clusters and offers a viable strategy for designing multicentered intermetalloids using small cluster building blocks, thus advancing both cluster chemistry and the development of functional intermetallic materials.

**Data availability statement**

All data that support the findings of this study are included within the article (and any supplementary files).

**Acknowledgements**

This study was supported by the National Natural Science Foundation of China (Grant Nos.12104393, 92461313, and 92161114), Funded by Science Research Project of Hebei Education Department (BJ2025076), the Innovation Capability Support Program of Shaanxi Province (No. 2023-CX-TD-49), and the Innovation Capability Improvement Project of Hebei Province (No. 22567605H).

**Conflict of interest**

The authors have declared that no competing interests exist.

**Reference**

[1] Kroto H W, Heath J R, O'Brien S C, et al. C$_{60}$: Buckminsterfullerene. *Nature*, 1985, 318: 162−163.
[2] Krätschmer W, Lamb L D, Fostiropoulos K, et al. Solid C$_{60}$: a new form of carbon. *Nature*, 1990, 347: 354−358.
[3] Moses M J, Fettinger J C, Eichhorn B W. Interpenetrating As$_{20}$ fullerene and Ni$_{12}$ icosahedra in the onion-skin [As@Ni$_{12}$As$_{20}$]$^{3−}$ ion. *Science*, 2003, 300: 778−780.
[4] Xu H L, Tkachenko N V, Szczepanik D W, et al. Symmetry collapse due to the presence of multiple local aromaticity in Ge$_{24}^{4−}$. *Nat. Commun.*, 2022, 13: 2149−2156.
[5] Yokoyama T, Nakajima A. Bridging the gas and condensed phases for metal-atom encapsulating silicon- and germanium-cage superatoms: Electrical properties of assembled superatoms. *Phys. Chem. Chem. Phys.*, 2023, 25: 9738−9752.
[6] Esenturk E N, Fettinger J, Lam Y F, et al. [Pt@Pb$_{12}$]$^{2−}$. *Angew. Chem. Int. Ed.*, 2004, 43: 2132−2134.
[7] Esenturk E N, Fettinger J, Eichhorn B. The Pb$_{12}^{2−}$ and Pb$_{10}^{2−}$ Zintl ions and the MPb$_{12}^{2−}$ and MPb$_{10}^{2−}$ cluster series where M = Ni, Pd, Pt. *J. Am. Chem. Soc.*, 2006, 128: 9178−9186.
[8] Wang J Q, Stegmaier S, Wahl B, et al. Step-by-step synthesis of the endohedral stannaspherene [Ir@Sn$_{12}$]$^{3−}$

[9] Zhou B, Krämer T, Thompson A L, et al. A highly distorted open-shell endohedral Zintl cluster: [Mn@Pb$_{12}$]$^{3-}$. *lnorg. Chem.*, 2011, 50: 8028−8037.

[10] Liu C, Li L J, Popov I A, et al. Symmetry reduction upon size mismatch: The non-icosahedral intermetalloid cluster [Co@Ge$_{12}$]$^{3-}$. *Chin. J. Chem.*, 2018, 36: 1165−1168.

[11] Mitzinger S, Broeckaert L, Massa W, et al. Understanding of multimetallic cluster growth. *Nat. Commun.*, 2016, 7: 10480−10489.

[12] Bandyopadhyay D, Sen P. Density functional investigation of structure and stability of Ge$_n$ and Ge$_n$Ni ($n$ = 1−20) clusters: Validity of the electron counting rule. *J. Phys. Chem. A*, 2010, 114: 1835−1842.

[13] Li X J, Su K H. Structure, stability and electronic property of the gold-doped germanium clusters: AuGe$_n$ ($n$ = 2−13). *Theor. Chem. Acc.*, 2009, 124: 345−354.

[14] Espinoza Quintero G, Duckworth J C A, Myers W K, et al. Synthesis and characterization of [Ru@Ge$_{12}$]$^{3-}$: An endohedral 3-connected cluster. *J. Am. Chem. Soc.*, 2014, 136: 1210−1213.

[15] Wang J, Han J G. Geometries and electronic properties of the tungsten-doped germanium clusters: WGe$_n$ ($n$ = 1−17). *J. Phys. Chem. A*, 2006, 110: 12670−12677.

[16] Khanna S N, Rao B K, Jena P. Magic numbers in metallo-inorganic clusters: Chromium encapsulated in silicon cages. *Phys. Rev. Lett.*, 2002, 89: 016803−016806.

[17] Fässler T F. The renaissance of homoatomic nine-atom polyhedra of the heavier carbon-group elements Si–Pb. *Coord. Chem. Rev.*, 2001, 215: 347−377.

[18] Fässler T F. Homoatomic polyhedra as structural modules in chemistry: What binds fullerenes and homonuclear Zintl ions? *Angew. Chem. Int. Ed.*, 2001, 40: 4161−4165.

[19] Rodríguez Kessler P L, Muñoz Castro A. Ligand-free supermolecules: [Pd$_2$@Ge$_{18}$]$^{4-}$ and [Pd$_2$@Sn$_{18}$]$^{4-}$ as multiple-bonded Zintl-ion clusters based on Pd@Ge$_9$ and Pd@Sn$_9$ assembled units. *Nanoscale*, 2024, 16: 5829−5835.

[20] Goicoechea J M, Sevov S C. [(Pd−Pd)@Ge$_{18}$]$^{4-}$: A palladium dimer inside the largest single-cage deltahedron. *J. Am. Chem. Soc.*, 2005, 127: 7676−7677.

[21] Liu C, Popov I A, Li L J, et al. [Co$_2$@Ge$_{16}$]$^{4-}$: Localized versus delocalized bonding in two isomeric intermetalloid clusters. *Chem. Eur. J.*, 2018, 24: 699−705.

[22] Xu H L, Qiao L, Sun Z M. [Co$_2$@(Ge$_{17}$Ni)]$^{4-}$: The first edge-sharing double-cage endohedral germanide. *Chem. Commun.*, 2022, 58: 3190−3193.

[23] Xu H L, Qiao L, Sun Z M. Orientational isomerism and its reactivity of a main group sandwich anion [Ge$_9$-In-Ge$_9$]$^{5-}$. *Chin. J. Chem.*, 2023, 41: 2432−2438.

[24] Wallach C, Selic Y, Witzel B J L, et al. Filled trivacant icosahedra as building fragments in 17-atom endohedral germanides [TM$_2$@Ge$_{17}$]$^{n-}$ (TM = Co, Ni). *Dalton Trans.*, 2021, 50: 13671−13675.

[25] Liu C, Jin X, Li L J, et al. Synthesis and structure of a family of rhodium polystannide clusters [Rh@Sn$_{10}$]$^{3-}$, [Rh@Sn$_{12}$]$^{3-}$, [Rh$_2$@Sn$_{17}$]$^{6-}$ and the first triply-fused stannide, [Rh$_3$@Sn$_{24}$]$^{5-}$. *Chem. Sci.*, 2019, 10: 4394−4401.

[26] Lips F, Clérac R, Dehnen S. [Pd$_3$Sn$_8$Bi$_6$]$^{4-}$: A 14-Vertex Sn/Bi cluster embedding a Pd$_3$ triangle. *J. Am. Chem. Soc.*, 2011, 133: 14168−14171.

[27] Xu H G, Zhang Z G, Feng Y, et al. Vanadium-doped small silicon clusters: Photoelectron spectroscopy and density-functional calculations. *Chem. Phys. Lett.*, 2010, 487: 204−208.

[28] Zhao J, Shi R, Linwei S, et al. Comprehensive genetic algorithm for ab initio global optimisation of clusters. *Mol. Simul.*, 2016, 42: 809−819.

[29] Delley B J. From molecules to solids with the DMol$^3$ approach. *J. Chem. Phys.*, 2000, 113: 7756−7764.

[30] Perdew J P, Burke K, Ernzerhof M. Generalized gradient approximation made simple. *Phys. Rev. Lett.*, 1996, 77: 3865−3868.

[31] Perdew J P, Burke K, Ernzerhof M. Generalized gradient approximation made simple [Phys. Rev. Lett. 77, 3865 (1996)]. *Phys. Rev. Lett.*, 1997, 78: 1396−1396.

[32] Wu X, Lu S J, Liang X, et al. Structures and electronic properties of $B_3Si_n^-$ ($n$ = 4–10) clusters: A combined ab initio and experimental study. *J. Chem. Phys.*, 2017, 146: 044306−044315.

[33] Wu X, Du Q, Zhou S, et al. Structures, stabilities and electronic properties of $Ti_mSi_n^-$ ($m$ = 1−2, $n$ = 14−20) clusters: a combined ab initio and experimental study. *Eur. Phys. J. Plus*, 2020, 135: 734−748.

[34] Liang X, Li X, Gao N, et al. Theoretical prediction for growth behavior and electronic properties of monoanionic $Ru_2Ge_n^-$ ($n$ = 3−20) clusters. *Inorg. Chim. Acta*, 2022, 542: 121141−121148.

[35] Liang X Q, Deng X J, Lu S J, et al. Probing structural, electronic, and magnetic properties of iron-doped semiconductor clusters $Fe_2Ge_n^{-/0}$ ($n$ = 3−12) via joint photoelectron spectroscopy and density functional study. *J. Phys. Chem. C*, 2017, 121: 7037−7046.

[36] Liang X, Kong X, Lu S J, et al. Structural evolution and magnetic properties of anionic clusters $Cr_2Ge_n$ ($n$ = 3−14): Photoelectron spectroscopy and density functional theory computation. *J. Phys. Condens. Matter*, 2018, 30: 335501−335511.

[37] Becke, Axel D. Density-functional thermochemistry. III. The role of exact exchange. *J. Chem. Phys.*, 1993, 98: 5648−5652.

[38] Frisch M J, Trucks G W, Schlegel H B, et al. Gaussian 16 Rev. A. 03. Wallingford, CT. 2016

[39] Silvi B, Savin A. Classification of chemical bonds based on topological analysis of electron localization functions. *Nature*, 1994, 371: 683−686.

[40] Savin A, Jepsen O, Flad J, et al. Electron iocalization in solid-state structures of the elements: The diamond structure. *Angew. Chem. Int. Ed.*, 1992, 31: 187−188.

[41] Lu T, Chen F W. Meaning and functional form of the electron localization function. *Acta Phys. Chim. Sin.*, 2011, 27: 2786−2792.

[42] Zubarev D Y, Boldyrev A I. Developing paradigms of chemical bonding: adaptive natural density partitioning. *Phys. Chem. Chem. Phys.*, 2008, 10: 5207−5217.

[43] Chen Z, Wannere C S, Corminboeuf C, et al. Nucleus-independent Chemical Shifts (NICS) as an aromaticity criterion. *Chem. Rev.*, 2005, 105: 3842−3888.

[44] Schleyer P v R, Maerker C, Dransfeld A, et al. Nucleus-independent chemical shifts: A simple and efficient aromaticity probe. *J. Am. Chem. Soc.*, 1996, 118: 6317−6318.

[45] Lu T, Chen F. Atomic dipole moment corrected hirshfeld population method. *J. Theor. Comput. Chem.*, 2012, 11: 163−183.

[46] Lu T, Chen F. Multiwfn: A multifunctional wavefunction analyzer. *J. Comput. Chem.*, 2012, 33: 580−592.

[47] Xie H H, Zhou L J, Cui P F, et al. Icosahedral carboranes-based metal nanoclusters: From fascinating structures to polychrome properties and applications. *Coord. Chem. Rev.*, 2025, 542: 216803−216820.

[48] Goicoechea J M, McGrady J E. On the structural landscape in endohedral silicon and germanium clusters, $M@Si_{12}$ and $M@Ge_{12}$. *Dalton Trans.*, 2015, 44: 6755−6766.

[49] Michalak A, Mitoraj M, Ziegler T. Bond orbitals from chemical valence theory. *J. Phys. Chem. A*, 2008, 112: 1933−1939.

[50] Cremer D, Kraka E. Chemical bonds without bonding electron density does the difference electron-density analysis suffice for a description of the chemical bond? *Angew. Chem. Int. Ed.*, 1984, 23: 627−628.

[51] Fliegl H, Taubert S, Lehtonen O, et al. The gauge including magnetically induced current method. *Phys. Chem. Chem. Phys.*, 2011, 13: 20500−20518.

[52] Steiner E, Fowler P W. Patterns of ring currents in conjugated molecules: A few-electron model based on orbital contributions. *J. Phys. Chem. A*, 2001, 105: 9553−9562.